# A comprehensive study on Blood cancer detection and classification using Convolutional neural network

Md Taimur Ahad
Associate Professor
Department of CSE
Daffodil International University
Dhaka, Bangladesh
taimurahad.cse@diu.edu.bd

Sajib Bin Mamun
Department of CSE
Daffodil International University
Dhaka, Bangladesh
sajib15-3435@diu.edu.bd

Sumaya Mustofa
Department of CSE
Daffodil International University
Dhaka, Bangladesh
sumaya15-3445@diu.edu.bd

Bo Song
Lecturer (Industrial Automation)
School of Engineering
University of Southern Queensland
Bo.Song@usq.edu.au

Yan Li
Professor (Computing)
School of Mathematics, Physics and Computing
Toowoomba Campus
University of Southern Queensland
Yan.Li@usq.edu.au

**Abstract** — Over the years in object detection several efficient Convolutional Neural Networks (CNN) networks, such as DenseNet201, InceptionV3, ResNet152v2, SEresNet152, VGG19, Xception gained significant attention due to their performance. Moreover, CNN paradigms have expanded to transfer learning and ensemble models from original CNN architectures. Research studies suggest that transfer learning and ensemble models are capable of increasing the accuracy of deep learning (DL) models. However, very few studies have conducted comprehensive experiments utilizing these techniques in detecting and localizing blood malignancies. Realizing the gap, this study conducted three experiments; in the first experiment- six original CNNs were used, in the second experiment - transfer learning and, in the third experiment a novel ensemble model DIX (DenseNet201, InceptionV3, and Xception) was developed to detect and classify blood cancer. The statistical result suggests that DIX outperformed the original and transfer learning performance, providing an accuracy of 99.12%. However, this study also provides a negative result in the case of transfer learning, as the transfer learning did not increase the accuracy of the original CNNs. Like many other cancers, blood cancer diseases require timely identification for effective treatment plans and increased survival possibilities. The high accuracy in detecting and categorization blood cancer detection using CNN suggests that the CNN model is promising in blood cancer disease detection. This research is significant in the fields of biomedical engineering, computer-aided disease diagnosis, and ML-based disease detection.

# 1. Introduction

In the deep learning architecture, Convolutional Neural Networks (CNN) is a type of architecture that depends on *'convolution'*, a mathematical combination of two functions that generates a third function, thereby connecting two sets of data. In contrast to Artificial Neural Networks (ANNs), which have a single layer, CNNs have a series of layers that are arranged in succession. To extract features from the input data, CNNs employ a convolutional layer, also known as a filter or kernel, which generates a feature map (Shah et al. 2023). The layering system of a CNN is made up of an input layer, several convolutional layers, pooling layers, a fully connected layer, and an output layer, in addition to hidden layers. CNNs are suitable for tasks that require the analysis of complex input data with spatial structure, such as image recognition and object detection. However, CNN has generated a lot of attention in data science since it has demonstrated the ability to locate, classify, and identify objects in images. Most importantly, CNN is useful in medical imaging because it can detect tumours and other irregularities more accurately in microscopic images, X-rays and MRI images.

CNN architectures can be divided into three broad categories, original, transfer learning, and ensemble technique. The first category, the original CNN architecture refers to a CNN network and algorithm that is available in Keras or Github. The second approach, transfer learning is based on the knowledge gained from a training dataset and is used for training a different but relevant task or field (Theodoris et al., 2023; Weiss et al., 2016). In this deep learning process, the first few layers are trained to define the characteristics of the task. The last few layers of the trained network can be removed and retrained with new layers for the target task. The last technique ensemble technique includes multiple CNNs and is expected to be more accurate than single CNN. By combining multiple models, an ensemble seeks to address the flaws of a single CNN and produce results (prediction and classification) based on the collective decision of participating CNNs in the ensemble model (Ahad et al., 2023; Baradaran Rezaei et al., 2023).

Blood cancer represents one of the most fatal cancers, approximately 1 in 6 deaths worldwide makes it the second leading cause of death globally. It accounts for approximately 9% of all cancers and is now ranked as the fourth most common cancer in both men and women worldwide (Hagar et al., 2023). The abnormal growth of cells in blood tissue causes blood cancer (Sajid et al. 2018). The blood cancer cell may vary in size, shape, and texture (Deepak et al. 2019; Yanming et al. 2022). These versatile characteristics of blood cell cancer allow CNN researchers to shift their focus toward applying CNN to develop models that can assist in detecting and classifying blood cancer (Hagar et al., 2023).

Realizing the effectiveness of CNN in the detection and classification of blood cancer, scholars such as Arkapravo and Mausumi (2022), Amjad et al. (2021), Saeed et al. (2024), Hareem et al. (2022). Hosseini et al., (2023), Rahman et al., (2023), and Wadhah et al. (2021) came forward and experimented with blood cancer detection and classification using CNN. These studies either customized the CNN or presented a new algorithm for blood cancer detection. Saeed et al., (2024) presented DeepLeukNet, a CNN-based model for acute lymphoblastic leukemia classification, Rahman et al., (2023) applied deep CNN with optimized features for multiclass blood cancer classification, Ahmed et al. (2023) experimented using hybrid techniques for the diagnosis of blood cancer, Devi et al., (2023) research applied segmentation and classification of white blood cancer cells from bone marrow microscopic images using duplet-convolutional neural network design.

Despite the significant advancement in computer-aided disease diagnosis, scholars such as Wang & Zhang (2020) warned of low accuracy and large false-positive values of ML. Accuracy in cancer detection and the classification of modalities are a concern as lower detection accuracy and large false-positive values will narrow the applicability and acceptability of CNN in cancer research (Hossain et al. 2023; Aladhadh et al. 2022). Identifying the correct type and grade of cancer in the early stages has an important role in the treatment plan (Shafique & Tehsin 2018). Sharma et. al. (2023) criticized, that though ensemble models have shown promise in improving blood cancer classification accuracy, the application of ensemble model blood cancer research is still relatively limited. Moreover, Chanda et. al., (2024) purported that the ensemble model is promising but the development of current ad-hoc developments overlooks redundant layers and suffers from imbalanced datasets and inadequate augmentation. Despite the fact data scientists have been making many efforts to utilize CNN, it is still not clear how existing CNN architectures perform in detecting blood cancer.

Following the gaps, this study aims to detect blood cancer using six original CNN architectures: DenseNet201, InceptionV3, ResNet152v2, SEresNet152, VGG19 and Xception. Moreover, we investigate if transfer learning can improve accuracy, and lastly, a hybrid ensemble model called DIX was developed, aiming to increase blood cancer detection accuracy.

However, the following are the primary contributions and novelties of this study:

1. This study compared the performance of six CNN networks, SE-ResNet152, MobileNetV2, VGG19, ResNet152v2, InceptionV3, and DenseNet201, as well as a transfer learning and ensemble model when analyzing images for blood cancer detection.
2. A novel DIX ensemble approach is introduced to remove the classification limitations of a singular CNN network. Experiment with three CNN models (Densenet121, InceptionV3, Xception) and then use a weighted voting-based ensemble approach. Multiple comparisons indicate that the DIX ensemble approach provides greater precision in this experiment.
3. The DIX ensemble model was also motivated by a study (Franc¸ois 2017) that questioned how much deep is necessary for cancer detection using CNN. Hu et al. (2018) in this regard advocated rather than deep or utilizing more convolution layers, the scholars suggested formulating CNN architecture using hyperparameter optimization, random search, and other more advanced model-based optimization techniques.

The paper is structured with a literature review, experimental setup, results of experiment, discussion, and conclusion. The results are published together with the experimental description because this study included three experiments for detection utilising such as six solo CNN networks (DenseNet201, InceptionV3, ResNet152v2, SEresNet152, VGG19, Xception), a transfer learning and ensemble model. Also provided are the study's limitations and future scopes.

## 2. Literature review

Various methods have been presented in the literature for the classification of blood cancer including CNN application, feature engineering, and classification.

Saeed et al., (2024) presented DeepLeukNet, a CNN-based leukemia classification. This research proposes an automated system for diagnosing Acute Lymphoblastic Leukemia disease using a CNN. For this purpose, simulation work has been performed over the Acute Lymphoblastic Leukemia-IDB 1 and Leukemia-lDB 2 datasets. Qualitative analysis has been performed by visualizing the intermediate layer activation, ConvNet filters and heatmap layers, and a comparative study has been performed with existing methods to validate the efficiency of the proposed model. The results showed that the proposed model attained 99.61% accuracy in Acute Lymphoblastic Leukemia diagnosis.

Ahmed et al., (2023) applied a hybrid technique for the diagnosis of acute lymphoblastic leukemia based on fusion of CNN features. The study applied the images of C-NMC 2019 and ALL-IDB2 datasets and then fed them to the active contour method to extract WBC-only regions for further analysis by three CNN models (DenseNet121, ResNet50, and MobileNet). The deep feature maps of DenseNet121-ResNet50, ResNet50-MobileNet, DenseNet121-MobileNet, and DenseNet121-ResNet50-MobileNet were merged and then classified by RF classifiers and XGBoost. The RF classifier with fused features for DenseNet121-ResNet50-MobileNet reached an AUC of 99.1%, accuracy of 98.8%, sensitivity of 98.45%, precision of 98.7%, and specificity of 98.85% for the C-NMC 2019 dataset.

Hosseini et al. (2023) provided a mobile application based on an efficient lightweight CNN model for the classification of blood cancer. Based on the well-designed and tuned model, a mobile application was designed for screening B-ALL from non-B-ALL cases. After comparing the efficiency of three notable architectures of lightweight CNN (EfficientNetB0, MobileNetV2, and NASNet Mobile), the most efficient model was selected, and the proposed model was accordingly configured and tuned. The proposed model achieved an accuracy of 100%.

According to Abir et al. (2022) proposed method utilizes several transfer learning models to classify lymphoblastic leukaemia, constituting an automated process. Additionally, the method explains the rationale behind each categorization by utilizing local interpretable model-agnostic explanations (LIME), ensuring validity and reliability. The InceptionV3 model achieved an accuracy of 98.38% using the suggested approach.

A new intelligent IoMT framework for the automatic classification of microscopic blood pictures was proposed by Karar et al. (2022). The framework comprises three primary stages. Firstly, wireless digital microscopy is employed to collect blood samples, which are then uploaded to a cloud server. Secondly, the cloud server utilizes a generative adversarial network (GAN) classifier to automatically identify blood conditions, such as healthy blood. Thirdly, the classification results are forwarded to a haematologist for medical approval. The developed GAN classifier was successfully evaluated on the ALL-IDB and ASH image bank, two open datasets. When compared to current state-of-the-art techniques which achieved the highest accuracy of 98.67% for binary classification (ALL or healthy) and 95.5% for multi-class classification (ALL, AML, and normal blood cells).

Sampathila et al. (2022) proposed a deep learning algorithm that utilizes microscopic images of blood smears as input data. The author developed a custom CNN model called ALLNET and trained it using open-source microscopic images of blood smears. After training and testing the model, the author's ALLNET model achieved impressive performance metrics, including a maximum accuracy of 95.54%, specificity of 95.81%, sensitivity of 95.91%, F1-score of 95.43%, and precision of 96%.

Tusar et al. (2022) model achieved an impressive accuracy of 98% in detecting multiple subtypes of ALL cells and also developed a telediagnosis software that can diagnose ALL subtypes in real-time using images from microscopic blood smears.

Jha et al. (2022) aimed to enhance leukaemia detection by employing deep ensemble learning on augmented datasets. The author utilized two different datasets from Kaggle and employed an artificial neural network for classification by ensemble classifier. Deep ensemble learning on enhanced augmented datasets was employed for more accurate identification of leukaemia cells, moving away from manual inspection and the proposed method achieved remarkable results, providing 100% accuracy with high-quality datasets and 96.3% accuracy even with poor-quality datasets.

An automatic classification technique of white blood cells (WBCs) using feature fusion techniques was proposed by Parayil et al. (2022). The author utilized various fusion techniques for feature extraction, including transfer-learning approaches such as DenseNet201 and VGG16. The study achieved an accuracy of 89.75% by combining feature fusion with a CNN.

Cheuque et al. (2022) applied a Faster R-CNN network to identify the region of interest (ROI) of WBCs and separate mononuclear cells from polymorphonuclear cells. After separation, two parallel CNNs with the MobileNet structure are used for recognizing subclasses within the identified WBCs at the second level. The author's Faster R-CNN and MobileNet CNNs are trained on a dataset of blood smear images containing annotations for the different WBC classes and the proposed model achieved a high-performance metric of around 98.4% across various evaluation metrics, including accuracy, recall, precision, and F1-score.

Sneha et al. (2022) introduced a technique for detecting acute lymphocytic leukaemia using a Chronological Sine Cosine Algorithm (SCA) based deep CNN. The model used Blood smear images, which are obtained from the Acute Lymphocytic Leukemia image database. To segment the images, A Mutual Information (MI) based hybrid model is proposed. This hybrid model combines the results of an Active Contour Model and Fuzzy C-means Algorithm (FCM) to accurately delineate regions of interest, likely leukemic cells, within the blood smear images. From the segmented images, statistical and textual features are extracted. Then the proposed methodology utilizes the SCA to optimize the weights of the deep CNN classifier. The optimized weights obtained from the SCA are applied to train the deep CNN classifier and the author's proposed model obtained an accuracy of 81%.

Baig et al. (2022) introduced a model that utilizes microscopic blood smear images to detect malignant leukaemia cells. The dataset, comprising approximately 4150 photos, was collected from a public directory. The images were regenerated in RGB colour space, multiplied with the source image, and resized to dimensions of [400, 400]. Six classification methods, including SVM, bagging ensemble, total boosts, RUSBoost, and fine KNN, were employed to evaluate the performance of feature extraction strategies. Among these, the bagging ensemble outperformed other classification algorithms, achieving the highest accuracy of 97.04%.

Khan et al. (2021) proposed a customized CNN model to classify four types of WBCs: neutrophils, eosinophils, monocytes, and lymphocytes, proposed by. The study includes pre-processing, segmentation, feature extraction, and others. In this study, a comprehensive evaluation was conducted using large training sets consisting of 9957 annotated blood smear images, along with test sets containing 2487 annotated images of WBCs (neutrophils, eosinophils, monocytes, and lymphocytes). Through a hierarchical learning process, the detection model achieved an impressive average accuracy of 98%. This thorough investigation demonstrated outstanding performance in accurately distinguishing between the four different types of WBCs.

Vogado et al. (2021) introduced LeukNet, a CNN designed for the classification of leukaemia based on microscopic images of blood cells. The authors LeukNet are inspired by the convolutional blocks of VGG-16 but integrate smaller dense layers. The parameters of LeukNet are defined through the evaluation of different CNN models and fine-tuning methods. Data augmentation operations are applied to expand the training dataset. The model's performance is evaluated using 5-fold cross-validation, resulting in an accuracy of 98.61%.

A CNN architecture presented by Claro et al. (2020) is capable of identifying blood slides with ALL, AML, and healthy blood slides (HBS). By utilizing 16 datasets comprising a total of 2,415 photos, the model achieved impressive accuracy and precision results of 97.18% and 97.23%, respectively. The performance of the proposed model was compared with state-of-the-art techniques, including those based on CNNs.

Vogado et al. (2018) focused on utilizing CNNs for the classification and diagnosis of leukaemia. A new database was constructed by combining three distinct databases from the literature, ensuring a comprehensive validation of the proposed methodology. Transfer learning was used to extract features from three state-of-the-art CNN architectures. These features were then selected based on their gain ratios and utilized as input to a Support Vector Machine (SVM) classifier and the author's proposed methodology achieved impressive hit rates exceeding 99%.

# 3. Research Methodology

The experiments for this study were conducted on Google CoLab. The purpose of using Google Colab is to take the opportunity to use TPU and GPU facilities. Keras library with TensorFlow was the chosen Python deep learning package for implementing the experiments of this study.

The selection of six CNNs was to cover all possible types of CNN as suggested by Asifullah et al. (2020). From spatial exploitation, VGG19, from Depth-based InceptionV3, ResNet152v2, from multipath DenseNet201, and feature mapping SEresNet152 were selected to cover an array of CNN architectures to study. The research methodology adopted in this study is presented in Figure 1:

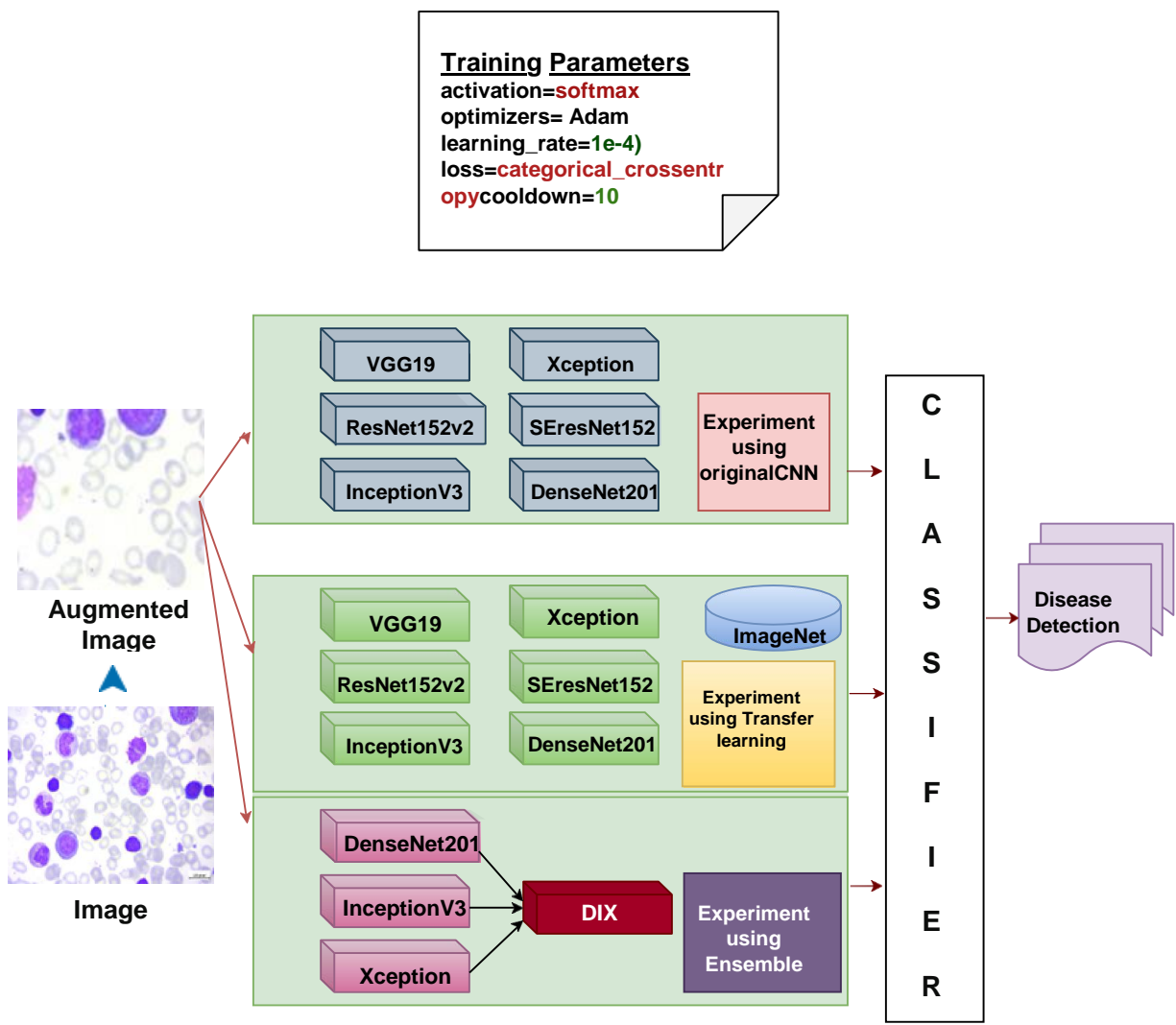

Figure 1. Diagram of the experiment.

### A. Datasets

The dataset for this study was collected from Kaggle. The dataset included 3235 microscopic images

(see Table 1). It was ensured that the dataset was almost balanced as there is criticism that blood cancer research utilized an imbalanced dataset (Chanda et. al., 2024). The dataset included: Malignant Early Pre-B, Malignant Pre-B, and Malignant Pro-B ALL. The images were in JPG format.

Medical images often suffer from low contrast and are subject to distortions caused by factors such as microscope sounds, image transmission, and digitization (Y Image 2020). Consequently, pre-processing of peripheral blood smear (PBS) images is essential to eliminate unwanted interferences present in raw PBS images.

Table 1. Distribution of peripheral blood smear (PBS) images used in the training, test and validation

|  | No of Images | Training images | Validation images |
| --- | --- | --- | --- |
| Benign | 505 | 353 | 101 |
| [Malignant] Pro-B | 796 | 557 | 159 |
| [Malignant] Pre-B | 955 | 668 | 191 |
| [Malignant] Early Pre-B | 979 | 685 | 195 |
| **Total** | **3235** | **2263** | **646** |

To separate training, testing, and validation, all raw images were divided into four classes. Figure 2 illustrates a sample dataset from the PBS.

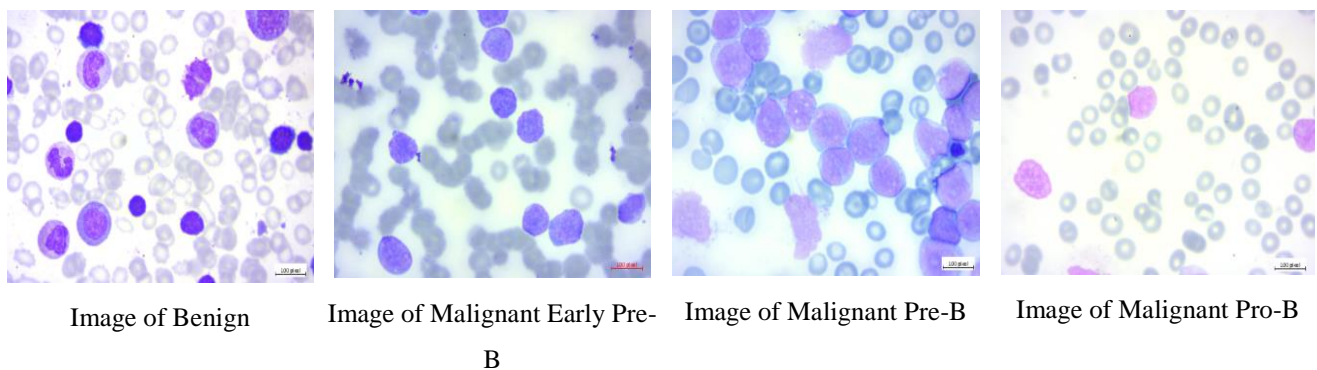

Image of Benign    Image of Malignant Early Pre-B    Image of Malignant Pre-B    Image of Malignant Pro-B

Figure 2. Example of 4 classes: Benign, Malignant Early Pre-B, Malignant Pre-B, and Malignant Pro-B.

### B. Image augmentation

During this phase, image augmentation was applied. The purpose of data augmentation was to make variations in data, improve the robustness of trained models to unfamiliar data, and increase model accuracy (Mohamed et al., 2021).

In this study, four data augmentation techniques: random cropping, horizontal flipping, vertical flipping, and centre cropping were applied to enhance the raw images. Furthermore, techniques such as Gaussian filtering, Linear Contrast adjustment, Median filtering, and Contrast Enhancement were utilized to improve contrast, reduce pixel and channel noise, eliminate bias fields, alter image colour, and enhance brightness (Abhijit et al., 2020). Figure 3, outlines the result of image augmentation for PBS blood cancer images.

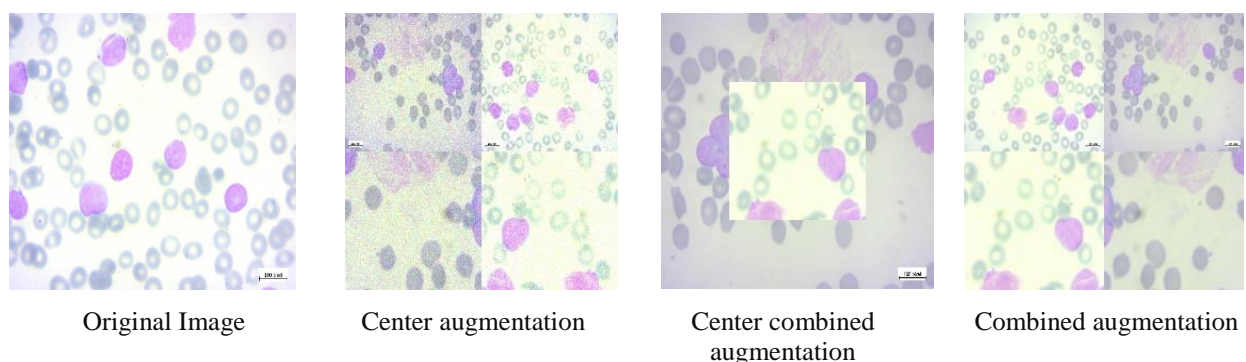

| Original Image | Center augmentation | Center combined augmentation | Combined augmentation |

Figure 3. Results of image augmentation.

### C. Training

In this stage, the CNN models such as VGG19, ResNet152v2, InceptionV3, Xception, SEresNet152, and DenseNet201 were trained using selected appropriate hyperparameters. Such as the diameters of convolutional windows, the number of network layers, and the number of filters in each layer. Three sets of hyperparameters were created, assessed, and applied to the CNN models in our study.

The models were trained with Early Stopping callbacks for 100 epochs, with patience of 10 iterations for all models. The patience threshold determines how many training epochs must pass without progress before training is stopped.

An Adam optimizer, SGD with momentum, and RMSProp were utilized to facilitate faster model training. Training times varied: InceptionV3 and Xception took 25 seconds per epoch, ResNet152v2 and VGG19 required 43 seconds per epoch, and DenseNet201 and SEresNet152 needed 55 seconds per epoch. The dataset used had no significant imbalances, so standard deviation was utilized as a measure of performance. Categorical cross-entropy served as the loss function for all architectures due to the focus on multi-class categorization. The final layers of the CNN models used SoftMax activation, with ReLU activation in intermediate layers. Hyperparameters included 60 epochs, 0.1 dropout rate, 1e-4 learning rate, and a batch size of 16. An Adam optimizer updated model weights, and images were resized to their respective architecture's default size. Figure 8 illustrates the process followed during the experiments.

## 4. Results of experiments

The experiment findings are presented in three parts, focusing on the original individual network, transfer learning, and ensemble methodologies, respectively. The outcomes aim to address the following research questions:

1. Which original CNN network provides better accuracy in detecting blood cancer?
2. Does transfer learning improve accuracy?
3. Does the ensemble technique improve the accuracy?

### A. Performance measurement

The performance of the experiments is evaluated using the following performance metrics:

Accuracy, defined as the proportion of correctly classified images to the total number of instances, is used for evaluating classification model performance. The accuracy equation is as follows:

$$\text{Accuracy} = \frac{TP+TN}{TP+FP+FN+TN} \qquad (i)$$

Precision indicates the proportion of accurately predicted cases that correspond to the positive class which is an important metric in situations where false positives (FP) are more concerning than false negatives (FN). It is mathematically expressed in the following equation:

$$\text{Precision} = \frac{TP}{TP+FP} \qquad (ii)$$

The recall represents the proportion of actual positive cases that our model was able to correctly capture. Recall is a useful statistic when false negatives (FN) are more concerning than false positives (FP). It is calculated using the following equation:

$$\text{Recall} = \frac{TP}{TP+FN} \qquad (iii)$$

F1-score is an additional metric for classification accuracy that considers both recall and precision. Since precision and recall are harmonic means, the F1 score achieves its highest value when Precision and Recall are balanced. It provides a comprehensive understanding of these two metrics. The equation is as follows:

$$F1\ score = \frac{2\times(Precision\times Recall)}{Precision+Recall} \qquad (iii)$$

This study answers its research questions in different sections that deal with each question separately.

## B. Experiment 1: Performance of the original CNN

This section presents the outcome of six original individual CNN networks: VGG19, ResNet152v2, InceptionV3, Xception, SEresNet152, and Densenet201. Initially, the categorization performance of these models is presented in Table 2. Subsequently, discuss the overall metrics for these models, highlighting their strengths and areas for improvement, and identifying contributing variables to the outcomes.

It's highlighted that the training models for all six architectures provided almost 100% accuracy. To enhance the input data, the GaussianBlur, LinearContrast, and AdditiveGaussianNoise algorithms were used for pre-processing. These techniques aim to improve contrast, eliminate noise from pixels and channels, adjust colour, and enhance image brightness.

Table 2. Training and model accuracy of six original CNN architectures.

| Architecture | Training Accuracy | Model Accuracy |
| --- | --- | --- |
| VGG19 | 94.84% | 96.94% |
| ResNet152v2 | 96.31% | 96.99% |
| InceptionV3 | 98.30% | 98.29% |
| Xception | 97.12% | 98.26% |
| SEresNet152 | 86.22% | 90.93% |
| DenseNet201 | 99.65% | 98.08% |

The accuracies shown in Table 3 represent the percentage of samples correctly identified among all samples. Regarding training accuracy, DenseNet201 achieved the highest value (99.65%), whereas SEresNet152 had the lowest accuracy (86.22%). In contrast, the model accuracies of DenseNet201, InceptionV3, and Xception were similar, with InceptionV3 achieving the highest percentage at 98.29%, while SEresNet had the lowest percentage at 90.93%. The performance of each model is shown individually in detail in Table 3.

Table 3. Precision, recall, f1, and support of six (6) results of original CNN networks (based on the number of images, n= numbers)

| | **VGG19** | | | |
|---|---|---|---|---|
| | Benign | Malignant Early Pre-B | Malignant Pre-B | Malignant Pro-B |
| Precision | 97% | 94% | 99% | 98% |
| Recall | 88% | 98% | 98% | 99% |
| F1-score | 93% | 96% | 99% | 99% |
| Support (N) | 1672 | 3254 | 3198 | 2628 |
| | **ResNet152v2** | | | |
| Precision | 96% | 95% | 98% | 99% |
| Recall | 93% | 97% | 99% | 100% |
| F1-score | 92% | 95% | 100% | 100% |
| Support (N) | 1641 | 3255 | 3196 | 2630 |
| | **InceptionV3** | | | |
| Precision | 99% | 96% | 100% | 100% |
| Recall | 90% | 100% | 100% | 100% |
| F1-score | 95% | 98% | 100% | 100% |
| Support (N) | 1672 | 3255 | 3197 | 2628 |
| | **Xception** | | | |
| Precision | 98% | 96% | 99% | 100% |
| Recall | 92% | 99% | 100% | 99% |
| F1-score | 95% | 98% | 100% | 100% |
| Support (N) | 1672 | 3252 | 3201 | 2627 |
| | **SEresNet152** | | | |
| Precision | 74% | 93% | 97% | 93% |
| Recall | 81% | 85% | 95% | 99% |
| F1-score | 77% | 89% | 96% | 96% |
| Support (N) | 1671 | 3253 | 3200 | 2628 |
| | **DenseNet201** | | | |
| Precision | 95% | 97% | 100% | 99% |
| Recall | 93% | 98% | 99% | 100% |
| F1-score | 94% | 97% | 100% | 100% |
| Support (N) | 1617 | 3255 | 3198 | 2629 |

Table 3 shows the Precision, Recall, F1-score, and Specificity obtained by the VGG19, ResNet152v2, InceptionV3, Xception, SEresNet152, and DenseNet-201 models for each class. After calculating the precision values for each class on the test dataset, the VGG19, ResNet152v2, InceptionV3, Xception, and DenseNet-201 architectures demonstrate superior performance. However, the SEresNet152 architecture provides poor performance, with the lowest identification.

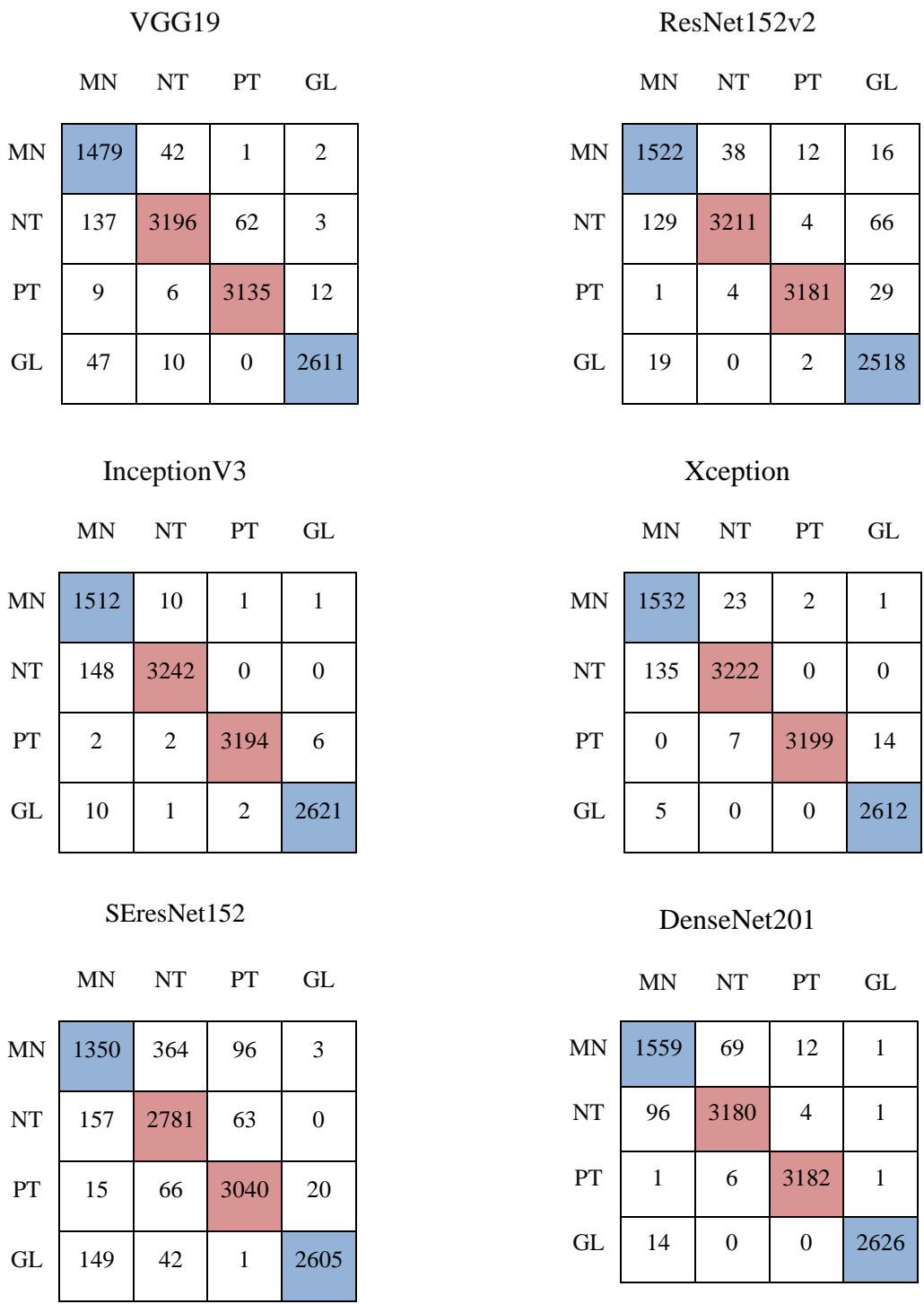

**MN = Benign**  
**NT = Malignant Early Pre-B**  
**PT = Malignant Pre-B**  
**GL = Malignant Pro-B**

Figure 3. Confusion matrix of six original CNN

The confusion matrix of the original (see Figure 3) reflects the results that Xception provides the highest true positive values (10565) than Inception (10500). However, Inception has fewer false positives and false negatives than Xception. Hence the model performance of Inception is slightly better than Xception.

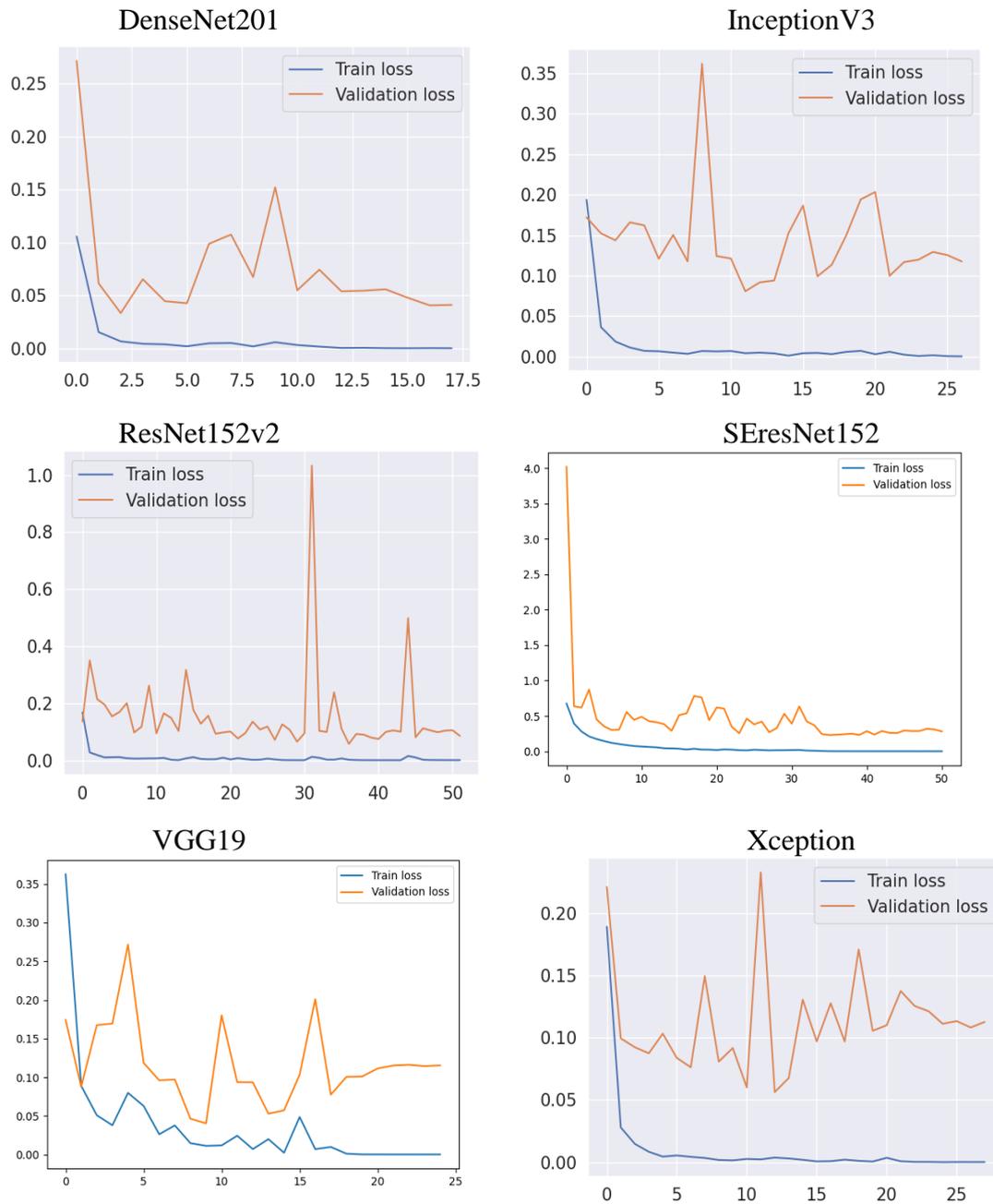

Figure 4. Loss curve of six original CNN.

Figures 4 and 5 show the loss and accuracy curves of the six original CNNs. Across all CNN models, it is observed that as the number of epochs increases, both training and validation loss decrease. While the loss lines show some variation as the number of epochs increases, they remain relatively constant afterwards. Additionally, there is no overfitting, and the training and validation data are clearly distinguished in the figure.

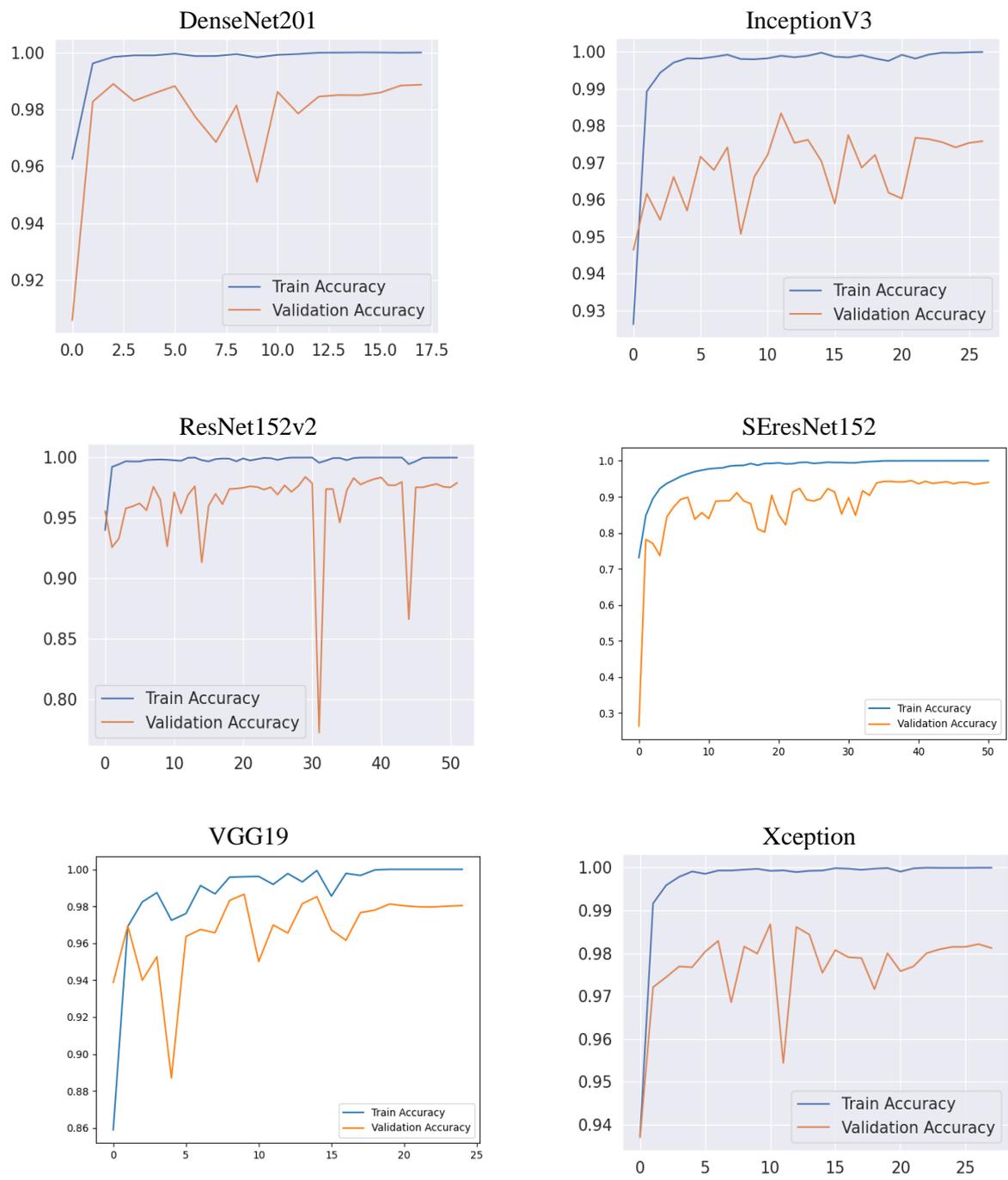

Figure 5. Accuracy curve of six original CNN.

## C. Experiment 2: Experimental process and result of transfer learning

This section presents the performance of six transfer learning CNN architectures: VGG19, ResNet152v2, InceptionV3, Xception, SEresNet152, and DenseNet201. Among these architectures,

DenseNet201, ResNet152v2, and Xception models performed well, while InceptionV3 models performed moderately, and VGG19 models performed poorly. The test accuracies shown in Table 5 were calculated using the ratio of correctly identified samples to all samples. Notably, the DenseNet201 model achieved the highest accuracy of 95%. However, it's important to note that the performance of the DenseNet201 network decreased from its original accuracy of 98.08% to 95.00% after transfer learning.

Table 5. Training and model accuracy of Transfer learning

| Architecture | Training Accuracy | Model Accuracy |
|---|---|---|
| VGG19 | 80.58% | 83.42% |
| ResNet152v2 | 88.79% | 90.89% |
| InceptionV3 | 86.37% | 91.41% |
| Xception | 88.14% | 93.55% |
| SEresNet152 | 91.94% | 94.16% |
| DenseNet201 | 92.57% | 95.00% |

Table 6 represents the Precision, Recall, F1-score, and Specificity obtained from CNN networks incorporating transfer learning. A model is considered exceptional if it shows high Precision, Recall, and Support. However, the experimental results show that VGG19 has a low precision in detecting blood cancer, with only 64%.

Table 6. Precision, Recall, F1, And Specificity Result of CNN Networks with Transfer Learning (N= Numbers)

| | VGG19 | | | |
|---|---|---|---|---|
| | Benign | Malignant Early Pre-B | Malignant Pre-B | Malignant Pro-B |
| Precision | 73% | 71% | 93% | 80% |
| Recall | 40% | 86% | 95% | 80% |
| F1-score | 52% | 78% | 94% | 80% |
| Support (N) | 1672 | 3255 | 3199 | 2626 |
| | ResNet152v2 | | | |
| Precision | 87% | 84% | 90% | 94% |
| Recall | 66% | 90% | 98% | 90% |
| F1-score | 75% | 87% | 94% | 92% |
| Support (N) | 1670 | 3253 | 3200 | 2629 |
| | InceptionV3 | | | |
| Precision | 73% | 71% | 93% | 80% |
| Recall | 40% | 86% | 95% | 80% |
| F1-score | 52% | 78% | 94% | 80% |
| Support (N) | 1672 | 3255 | 3199 | 2626 |
| | Xception | | | |
| Precision | 73% | 71% | 93% | 80% |
| Recall | 40% | 86% | 95% | 80% |
| F1-score | 52% | 78% | 94% | 80% |
| Support (N) | 1672 | 3255 | 3199 | 2626 |
| | SEresNet152 | | | |
| Precision | 94% | 87% | 93% | 96% |
| Recall | 70% | 93% | 100% | 94% |
| F1-score | 80% | 90% | 96% | 95% |
| Support (N) | 1670 | 3255 | 3199 | 2628 |
| | Densenet201 | | | |
| Precision | 94% | 88% | 94% | 97% |
| Recall | 71% | 93% | 99% | 97% |
| F1-score | 80% | 91% | 96% | 97% |
| Support (N) | 1617 | 3256 | 3200 | 2625 |

**MN = Benign    NT = Malignant Early Pre-B    PT= Malignant Pre-B    GL = Malignant Pro-B**

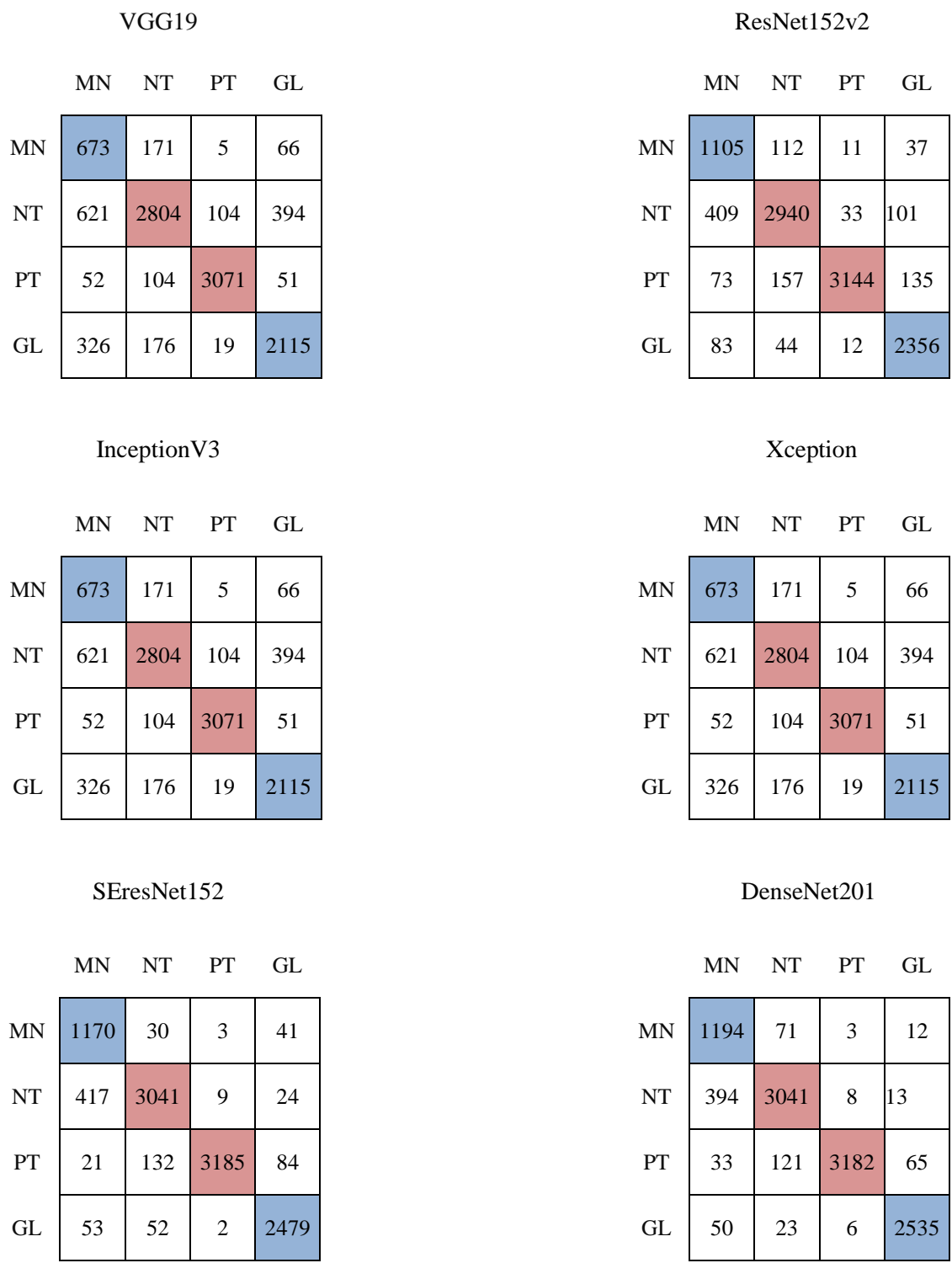

Figure 6. Confusion Matrix of Transfer Learning CNN.

The confusion matrix of the transfer learning (see Figure 6) reflects the results that DenseNet201 provides the highest true positive values (9875), ResNet152v2 has the second highest true positive values (9545) and other models have the same amount of true positive value (8663). Hence the model performance of DenseNet201 is slightly better than the other transfer learning models.

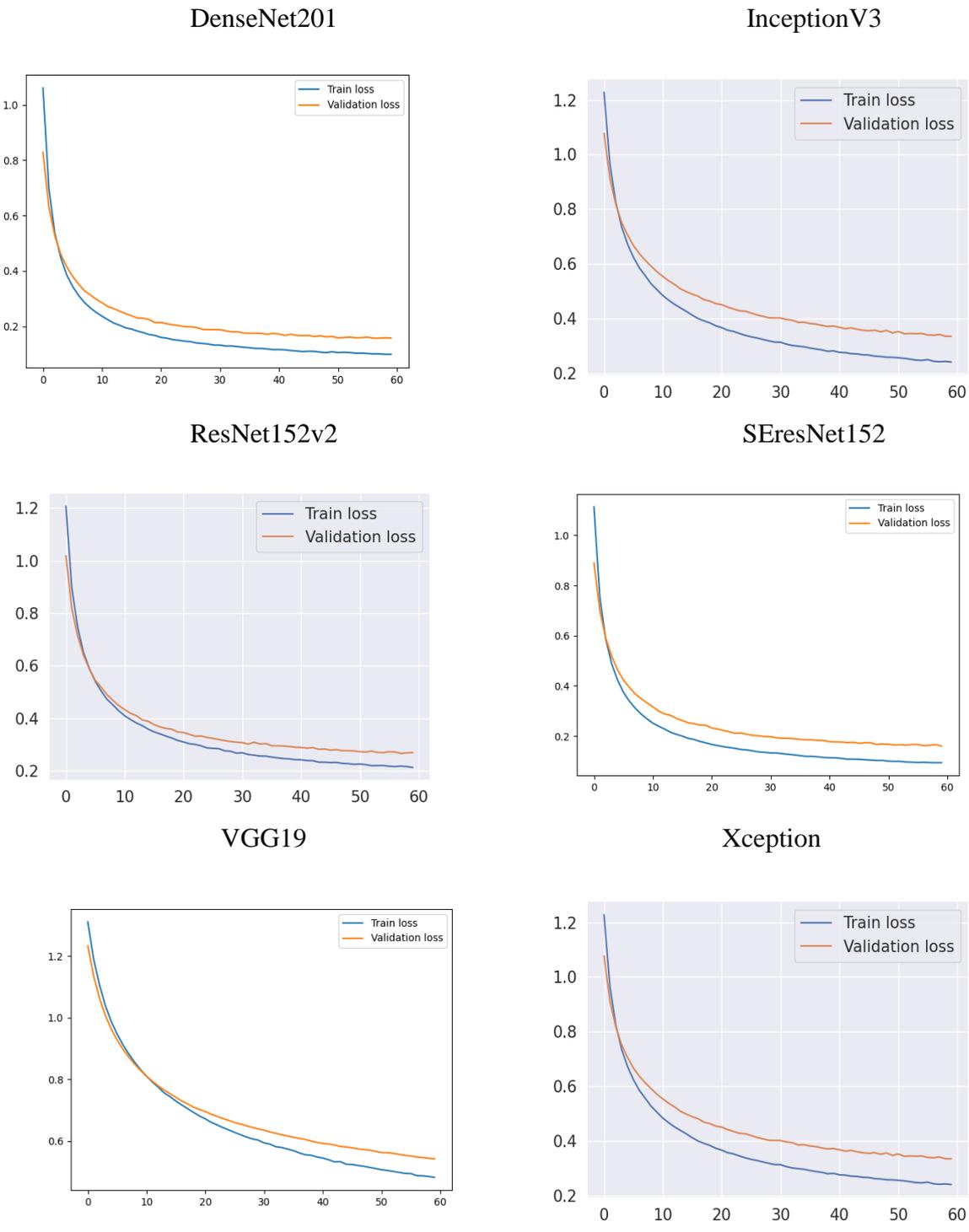

Figure 7. Loss curve of Transfer learning CNN

Figure 7 shows the loss curves of the six transfer learning models. Across all transfer learning models, it is observed that as the number of epochs increases, both training and validation loss decrease.

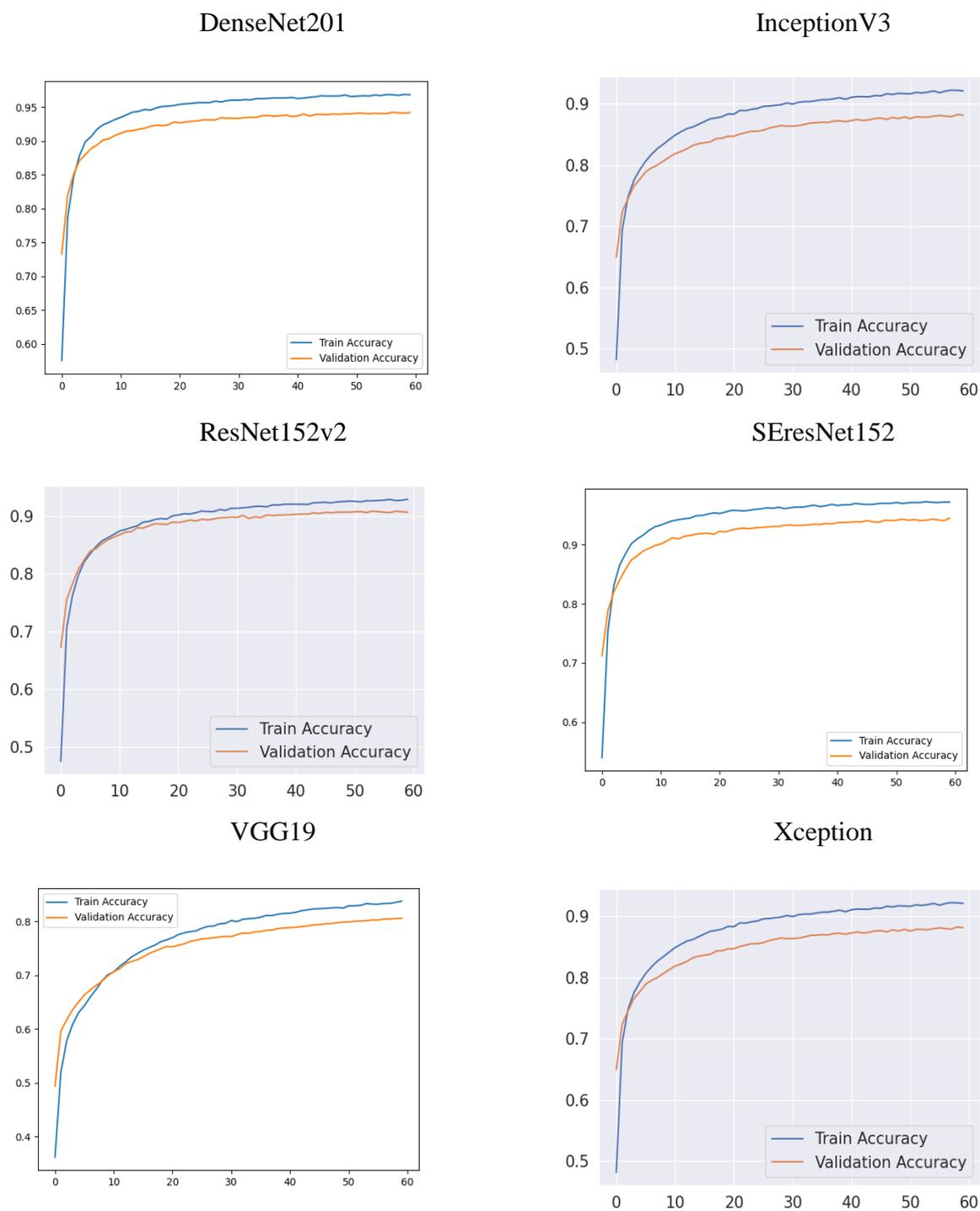

Figure 8. Accuracy curve of Transfer learning CNN

Figure 8 shows the accuracy curves of six transfer learning models. Across all networks, it is observed that as the number of epochs increases, both train and validation accuracy significantly increase. While the graphs show slight changes in the accuracy lines with increasing epochs, they remain relatively stable thereafter.

**D. Experiment 3: Experimental process and the result of the Ensemble model**

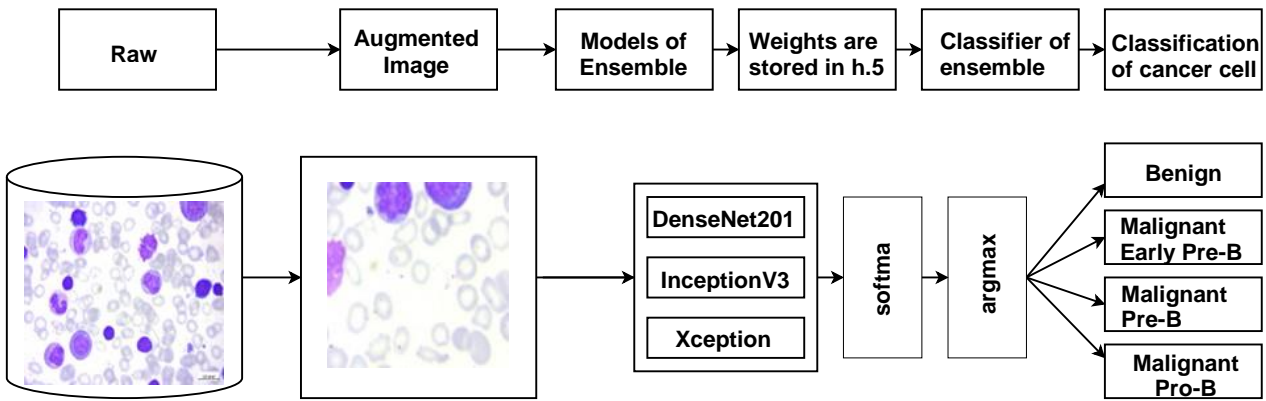

Figure 9. Block diagram of Ensemble DIX model

In this study, an ensemble model was created comprising three distinct original CNN models: DenseNet201, InceptionV3, and Xception. To enhance the training process, this study used a transfer learning approach but transfer learning performance was not as expected. The output from these models was then passed through a post-processing block, which included a layer specific to each model, a pass, and a final logit layer for image classification. All models were trained for 60 epochs with Early Stopping (56 epochs) callbacks and patience of 15 epochs. The Adam optimizer, incorporating SGD with momentum and RMSProp, was employed with specific learning rate parameters to achieve faster convergence. The same optimizer was applied to all three models, and the models were subsequently saved as *.h5* files. Each epoch of the DIX (DenseNet201, InceptionV3, Xception) model training lasted 68 seconds.

**E. Model Development**

In this study, one of the main goals was to develop an ensemble model that improves the accuracy of detection and classification of blood cancer. The main reason behind the development of the ensemble model is even if a weak classifier got a wrong prediction, the whole ensemble classifier (strong classifier) could correct the error back yet. In addition, the ensemble method could reduce the variance. In this experiment, DenseNet201, InceptionV3, and Xception were selected as the candidates for the ensemble model 'DIX' (see Table 3). The purpose was to mix a strong classifier with a weak classifier to validate the capabilities of the ensemble as suggested by Sharma et. al. (2023).

The ensemble model used in this study aggregates the Sum of Probability values from three CNNs (DenseNet201, InceptionV3, Xception) and calculates the sum of probabilities for each class from the individual CNN architectures. The final prediction is determined by considering the class with the maximum normalized sum.

The ensemble function $\mathcal{E}(\cdot) f\ n\ c$ is used, where $n = 3$ in our proposed framework. Each class receives $n$ confidence values for a given image $I$.

The final classification decision is made based on the maximum likelihood among the classes. The confidence values $s_{ij}$, where $i \in \{1, 2, …, m\}$ and $j \in \{1, 2, …, n\}$, are aggregated using the Sum of Probabilities.

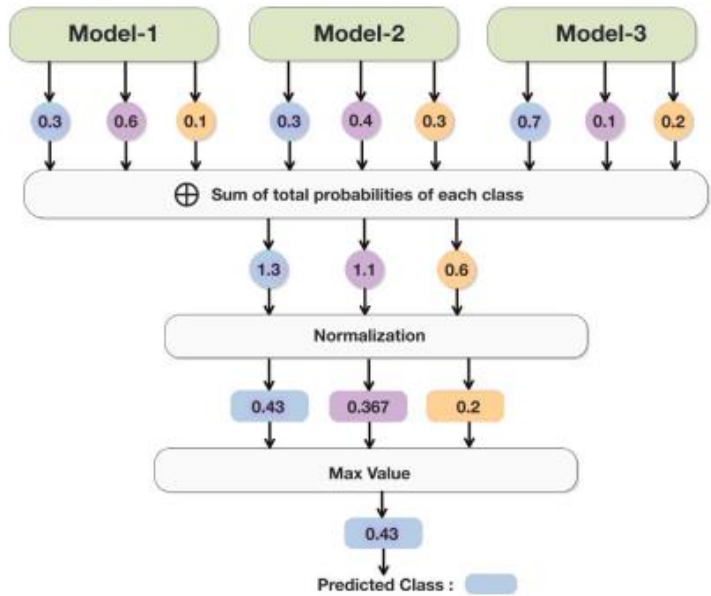

Figure 10. Creation of the proposed DIX ensemble model.

In Figure. 10, the individual summation of prediction values is shown as 1.3, 1.1, and 0.6 for class-1, class-2, and class-3, respectively. The normalization step ensures that the probability values sum to one, and the class with the largest normalized sum is selected as the predicted class. The SoP aggregation process for the purposed network is formulated in Eq. (v), where $i$ is the index for the class and $j$ is the index for the Deep Convolutional Neural Network (DCNN) models. Thus, $s_{ij}$ means the prediction value of $i$th class out of $m$ number of classes and $j$th model out of $n$ number of models.

A normalization factor $\sum_{i=1}^{m}\sum_{j=1}^{n} s_{ij}$ is used to normalize the values after the summation of corresponding class values $\sum_{j=1}^{n} s_{ij}$

$$S_{pred} = \max\left(\frac{\sum_{j=1}^{n} s_{ij}}{\sum_{i=1}^{m}\sum_{j=1}^{n} s_{ij}}, \forall_i\right) \quad (v)$$

---

**Algorithm 1** Ensemble procedure.

---

1: Input: [DenseNet201, InceptionV3, Xception], test_dataset
2: Output: ensemble_prediction
3: **models** ← [DenseNet201, InceptionV3, Xception]
4: for all *model in models* do
5: **predictions** ← model_predict (*test_dataset*)
6: end for
7: **pred_array** ← array (*predictions*)
8: **pred_sum** ← sum (*pred_array, axis* = 0)
9: *ensemble_pred* ← argmax (*pred_sum, axis* = 1)
10: **ensemble_prediction** ← *ensemble_pre*d

---

Table 7. Training and model accuracy of Ensemble model DIX (DenseNet201, VGG19 and SEresNet152)

| Architecture | Training Accuracy | Model Accuracy |
|---|---|---|
| DenseNet-201, InceptionV3 and Xception (DIX) | 97.63% | 99.12% |

Table 8. Precision, Recall, F1-Score and Support of Ensemble model DIX

| | **DIX** | | | |
|---|---|---|---|---|
| | Benign | Malignant Early Pre-B | Malignant Pre-B | Malignant Pro-B |
| precision | 100% | 95% | 100% | 100% |
| Recall | 88% | 100% | 100% | 100% |
| f1-score | 93% | 97% | 99% | 100% |
| support (N) | 1672 | 3254 | 3198 | 2628 |

In precision, the ensemble algorithm achieved 100% accuracy on Benign, Malignant Pre-B ALL, and Malignant Pro-B ALL, Outperforming both transfer learning and the original CNN model. However, the lowest accuracy for meningioma was observed in the F1-score. Table 8 presents the Precision,

Recall, F1-Score, and Specificity of the Ensemble model DIX when employing the ensemble technique.

**MN = Benign       NT = Malignant Early Pre-B       PT= Malignant Pre-B     GL = Malignant Pro-B**

|    | MN   | NT   | PT   | GL   |
|----|------|------|------|------|
| MN | 1479 | 3    | 0    | 0    |
| NT | 161  | 3248 | 0    | 2    |
| PT | 26   | 3    | 3198 | 2    |
| GL | 6    | 0    | 0    | 2624 |

Figure 11: Confusion matrix of Ensemble model DIX.

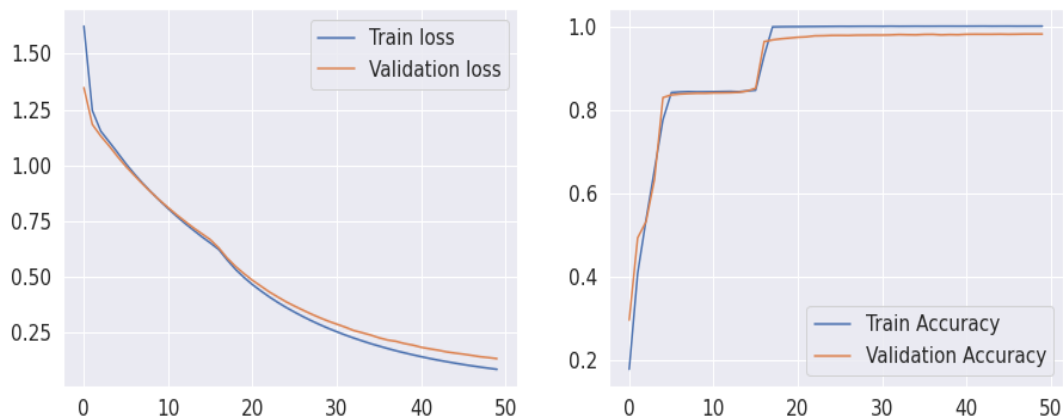

Figure 12. Loss and Accuracy curve of Ensemble model DIX.

The ensemble model outperformed the original CNN architecture with a 99.12% accuracy rate. If any model provides less accuracy, the ensemble model is utilized to improve model accuracy and performance. Additionally, the ensemble DIX model gives 0.83% better performance than the original CNN architecture and 4.12% better performance than transfer learning. Table 8 presents the Precision, Recall, F1-score, and Specificity results of the CNN networks using the ensemble. Figure 12 illustrates the training accuracy and validation accuracy of the ensemble model using data from DenseNet201,

InceptionV3, and Xception. Figure 12 illustrates that there is no overfitting, and the training and validation data are appropriately separated.

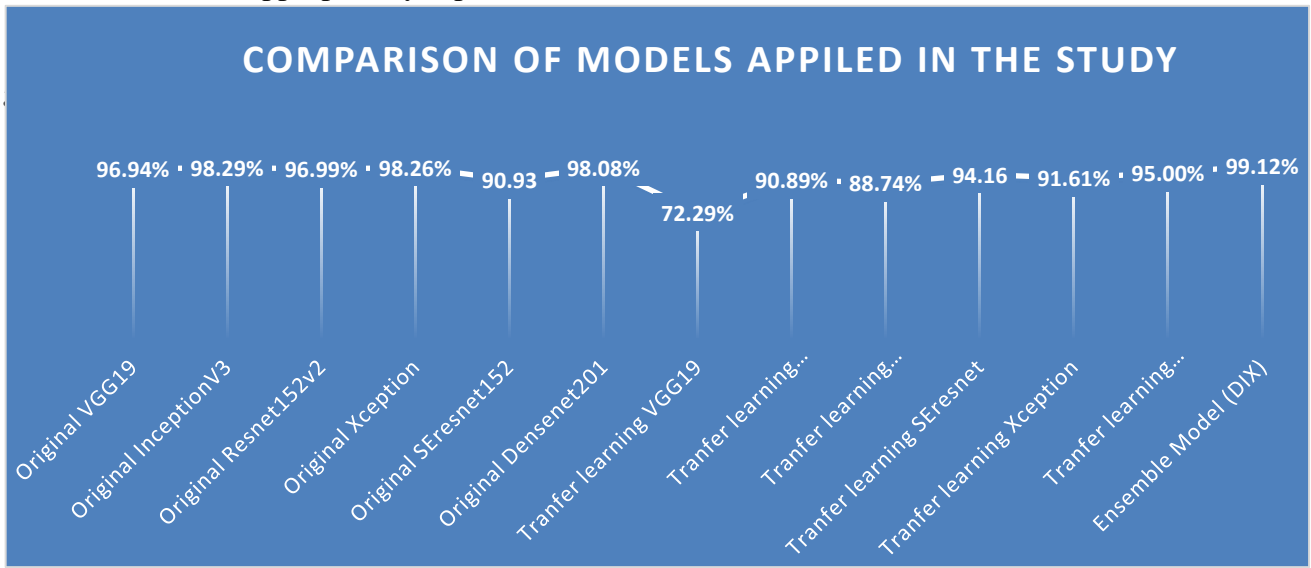

Figure 13. Accuracy comparison among individual CNN and transfer learning

In this research, a comprehensive examination of CNN's performance in detecting and classifying four types of blood cancer images is conducted using 3235 images. Six different CNN models: VGG19, ResNet152v2, InceptionV3, Xception, SEResNet152, and DenseNet201 were applied to four classes of blood cancer (see Figure 13). DenseNet201, InceptionV3, and Xception were among the original individual networks that provided the best classification outcomes for diagnosing blood cancer. Particularly, InceptionV3 achieved the highest accuracy (98.29%) among the original CNN models. Though significant studies purported that transfer learning may increase the accuracy, however, in our case the transfer learning approach did not increase the accuracy but rather declined that the CNN networks. The result aligns with previous studies (Ahad et al. 2023; Krishnaswamy et al. 2020; Jayme et al. 2018), indicating that accuracy may decrease when the input image differs significantly from the trained data in the ImageNet Dataset. Performance was impacted by background noise and the application of various augmentation strategies independently with the test sets. In the case of the original CNN, the models were trained and evaluated using comparable input, enhancing its prediction capabilities for unseen data; however, in the transfer learning, the test datasets likely affected its capacity to predict.

The ensemble model, including three CNNs (DenseNet201, InceptionV3, and Xception) provides an impressive accuracy of 99.12% outperforming the original CNN architecture (DenseNet201, InceptionV3, and Xception). Moreover, it achieved a 0.83% improvement over the original CNN architecture. As expected from our analysis, the combination of deep learning models outperformed a single CNN architecture in terms of accuracy.

## 6. Inference of the study

This study investigates deep learning techniques for the detection and classification of blood cancer. Key conclusions drawn from this study are as follows:

1. The classification accuracy differs while using the InceptionV3, Xception, ResNet152v2, SEresnet152, DenseNet201 and VGG19 models for the same microscopic image of blood cancer and the same training set of data. Experimented results show that 98.29%, 98.26%, 96.99%, 90.93%, 98.08%, and 96.94% accuracy, respectively (See the Figure 13). Inception-v3 has 48 deep convolutional layers. The ImageNet database contains a pre-trained version of the network that has been trained on more than a million photos. Additionally, Xception is a 71-layer CNN. We have contrasted the Inception-v3 and Xception architectures' accuracy. Both offer a higher level of microscopic picture precision based on the comparison at epoch 60 as a sequential method. VGG-19 is a 19-layer CNN. ResNet152 can have a very deep network with up to 152 layers by learning the residual representation functions rather than the signal representation directly. The accuracy of the VGG19 and ResNet152 architectures has been compared, even though both of these models focus on the same image categorization issue. We have determined that the ResNet is a superior architectural design. For microscopic images, however, SEresNet152 performs worse than ResNet. As a generalized feature extractor that is built as a ResNet variation and moved to the destination dataset. The DenseNet201 (Dense Convolutional Network) design seeks to enhance the depth of deep learning networks while simultaneously enhancing training effectiveness by using shorter connections between the layers. Because of the gap between the input and output layers in high-level neural networks, DenseNet201 was created specifically to boost the accuracy caused by this phenomenon.

2. However, analysis of this study identifies instances of negative transfer when there are significant disparities between the target and source datasets. While transfer learning has the potential to effectively train deep learning models, our findings highlight its limitations in scenarios where datasets differ substantially.

3. The ensemble of numerous models may still perform better than one model, even in the case where a given CNN architecture fails. This paper proposes an ensemble model (DIX) with the highest accuracy of 99.12% using DenseNet201, Inception-v3, and Xception. Using DIX (DenseNet201, Inception-v3, and Xception) turned into a generalized feature extractor of the target dataset with a deep network with up to 71 layers, an ensemble method is used to keep the connections between the layers simple.

# 7. Contributions

To better understand the performance of CNNs in blood cancer research, this research conducted a comparative study on blood cancer detection and classification. Utilizing four classes and 3235 images, the classification results using original CNNs, transfer learning, and ensemble learning approaches have been evaluated. Performance metrics such as inference time, model complexity, computational complexity, and mean-per-class accuracy were considered across various CNN designs. An ensemble model was developed to enhance accuracy, demonstrating superior performance compared to individual CNN architectures. The ensemble model DIX (composed of DenseNet201, InceptionV3, and Xception) achieved an accuracy of 99.12%, indicating its ability to classify blood cancer more accurately. Therefore, the DIX ensemble model shows promise for enabling a more precise diagnosis of blood cancer.

# 8. Conclusion and future research direction

This research is an effort to establish that CNN-based models can effectively identify blood cancer, increase accuracy, and reduce false-positive and false-negative values of the dataset. Using balanced data, appropriate augmentation, and applying hyperparameters techniques, this study provides a comprehensive evaluation of six CNN, transfer learning, and ensemble models in detecting and categorizing blood cancer. The findings of this paper suggest that CNN-based detection methods are effective even with small datasets, addressing a gap in specific research on blood cancer identification. Comparing different CNN models, including transfer learning and ensemble methods, it was found that the ensemble DIX model, combining DenseNet201, InceptionV3, and Xception, achieves the highest accuracy in classifying blood cancer.

This research also offers some improvement opportunities. The application of Adadelta, FTRL, NAdam, Adadelta, and other optimizers can be tested. Since Google Colab provides service for a limited time, in the future hyperparameter tuning, base model training on databases other than Imagenet (this research used Imagenet as the base database in transfer learning), and more ensemble models can be developed. Another drawback of the study is that secondary data from the public domain, rather than primary data gathered directly from the hospital. Future research can include data from hospitals and qualitative experiments can be employed in association with the medical expert to confirm if the result is accurate from the medical science perspective.

The experiments are not yet ready to use for end users, the experiments are only from a data-science perspective. To make the research from a user perspective, in future, the model can be deployed for mobile or portable device-oriented.

In the future more ensemble techniques can be adapted to prove the applicability of the ensemble model, and more research can be improved to improve the performance of transfer learning can be adopted. However, this research highlights the potential of CNN and its variants in improving blood cancer detection and classification.

**Declaration of competing interest**

The authors declare that they have no known competing financial interests or personal relationships that could have appeared to influence the work reported in this paper.**Data availability**

The data of this research are stored in the Kaggle respiratory.

**Funding Statement**

The work was not supported by any funding and neither did any of the researchers receive funds.